\documentclass[aps,prl,twocolumn,showpacs,superscriptaddress]{revtex4}
\usepackage{graphicx}
\begin{document}
\title{Renormalization Group Limit Cycles in Quantum Mechanical Problems}
\author{Erich J. Mueller}
\affiliation{ Laboratory of Atomic and Solid State Physics, Cornell University, Ithaca, New York 14853}
%\email{em256@cornell.edu}
\author{Tin-Lun Ho}
\affiliation{Department of Physics,  The Ohio State University, Columbus, Ohio
43210}
%\email{ho@mps.ohio-state.edu}

\newcommand{\sgn}{\mathrm{sgn}}

\begin{abstract}
We formulate a renormalization group (RG) for the interaction parameters of the general two-body problem and show how a limit cycle emerges in the RG flow if the interaction 
approaches an inverse square law.  This limit cycle generates a scaling structure in the energy spectrum.  
Our demonstration is relevant to the Efimov problem in nuclear physics where similar scaling appears. 
\end{abstract}
\date{Mar 10, 2004}
\maketitle

In his famous 1971 paper on the Renormalization Group (RG) \cite{Wilson71}, Kenneth Wilson pointed out that renormalization group equations (which describe how coupling constants evolve as one rescales a system) can display all of the rich structure of dynamical systems including fixed points, chaos, and limit cycles.  Yet nearly all examples of RG flows in statistical physics contain only fixed points.  Aside from Huse's 1981 three state chiral model \cite{Huse}, all of the exceptions were found in the last few years.  These include discrete models of
Glazek and Wilson \cite{GW} and 
field theory models of LeClair et al. \cite{leclair}. Amid these revelations, Albeverio et al. \cite{Albe-81} realized that the Efimov effect \cite{Efimov} in nuclear physics is associated with a renomalization group limit cycle.   %The cycles in the Efimov problem are closely related to the behavior of atoms in an inverse square potential \cite{inversesquare}.
 Here we continue this line of inquiry and explore limit cycles in the RG flow of potential scattering problems (whose origin and structure coincide with those of the Efimov problem \cite{Efimov}).

The Efimov effect  occurs in a system of three particles, when the pairwise interaction is at resonance (i.e. the two-body potential is ``tuned" to the point where a bound state appears just below the continuum).  Under these circumstances, the three-body system has infinitely many bound states, whose energies form a geometric series.  The ratio between successive binding energies is a universal number.  Since the deuteron (a bound state of a neutron and proton) has such a small binding energy, the proton-neutron system is near resonance and the Efimov problem is relevant to nuclear physics \cite{Efimov,Braaten}.  
%For example, Braaten \cite{Braaten} argues that the Efimov effect is responsible for the ground state and excitation energies of a triton.  
On the other hand, since there are no ways to tune the nuclear potential, exact resonances do not
occur in nuclear systems, frustrating studies of the Efimov effect.  

The 1996 discovery of Bose-Einstein condensation in atomic gases \cite{BEC} has changed this situation.  Soon after the discovery, Ketterle \cite{Ketterle} used a magnetic field to tune the atomic scattering to resonance, creating a new system for studying the Efimov effect.

The Efimov problem can be mapped onto a quantum scattering problem.  When the pairwise interactions are resonant, the important degree of freedom for the three-body problem is the {\em hyper-spherical radius} $R$: $R^2={\sum_{i\neq j}|{\bf r_i-r_j}|^2}$, where  $\bf r_j$ is the position of the $j$'th particle.  In this variable, the wavefunction obeys a Schrodinger equation with a $1/R^2$ potential \cite{Efimov}.  Since $1/R^2$ scales as $\nabla^2$, the quantum mechanical problem is scale invariant and (as we verify below) can have a RG limit cycle.  

In this paper we formulate a renormalization group for the interaction constants of 
the two-particle quantum mechanical scattering problem and show how limit cycle and fixed point behaviors emerge as the interaction approaches and deviates from scale invariance.  Although two-particle problems are exactly solvable, framing the problem in the RG language gives insights which are not readily extracted from the exact solutions.  For example, our calculation provides a
 transparent connection between scale invariance and limit cycle behavior. We believe our simple 
derivation will be helpful in understanding the generality of the limit cycles in the RG. 

Several recent papers have used RG ideas to discuss potential scattering \cite{barford,beane} and used these ideas to investigate limit cycles \cite{bawin}.
These authors were interested in producing low energy effective field theories (EFT) and in constructing short-range regularizations to keep energy eigenstates constant. 
%Motivated by attempts to understand few-body and many-body problems in nuclear and hadronic physics, the effective field theory community has recently been producing RG approaches to understanding quantum mechanical problems \cite{barford,beane,bawin,birse}.  Several of these even explore the limit-cycle structure of the RG flows of an inverse square potential \cite{beane}.
Here, we are interested in the global behavior of the RG  flow of 
the entire set of interaction constants as the system is scaled to large and short distances. Our method is therefore capable of exploring both high and low energy behaviors.

\noindent
{\bf The quantum mechanical two-body problem}: The two-body problem in the center of mass frame has the form of a single particle
scattering problem in a central potential $V(r)$, 
where $r$ is the distance between the particles \cite{qmtext}.  
The energy eigenfunctions $\psi({\bf r},E)$ obey the Schrodinger equation $(- (\hbar^2/M)\nabla +V(r))\psi({\bf r} ,E) = E\psi({\bf r},E)$, where $M/2$ is the reduced mass.
For our discussion, we solely
consider s-wave scattering, so $\psi$ is only a function of $r=|{\bf r}|$ and in $d$ dimensions, $\nabla^2 = \partial_{r}^2 + (d-1)r^{-1}\partial_{r} $;  higher partial 
waves are treated with similar methods. 

The goal of the RG is to analyze a system on different length scale.  It is therefore necessary to
divide the potential into long and short range parts \cite{desire}.  We introduce a cutoff $r_0$, and let $V(r)$ for $r>r_0$ specify the long distance properties.  Short range structure is naturally encoded in a boundary condition at $r_0$.  The most general boundary condition consists of specifying the logarithmic derivative $g(r,E)=r \psi'(r,E)/\psi(r,E)$ at $r=r_0$ for all $E$.  We express length and energy in dimensionless form, $x = r/r_{o}$ ($0\leq x< +\infty$), ${\cal E} = E/\varepsilon$, and $U=V/\varepsilon$, where $\varepsilon=\hbar^2/Mr^{2}_{o}$.  The same symbol will be used for functions of the scaled and original variables [for example, $\psi(x,{\cal E}) = \psi(r,E)$, and $g(x,{\cal E}) = x\psi'(x,{\cal E})/\psi(x,{\cal E})= g(r,E)$].  The two-body problem is then specified by
 \begin{eqnarray}
 [-\nabla^2 + U(x)] \psi(x,{\cal E})  &=&  {\cal E}\psi(x, {\cal E}),\quad x>1 \nonumber \\
 \left[x\psi'(x,{\cal E})/\psi(x,{\cal E})\right]_{x=1} &=& g(1,{\cal E}).   
 \label{I} \end{eqnarray}
Since the choice of $r_{o}$ is arbitrary, we can choose a different point, say, $\lambda r_{o}$ to impose the boundary condition,  where $\lambda$ is a scale factor.  The corresponding equations are\begin{eqnarray}
 [-\nabla^2 + U(x)] \psi(x,{\cal E})  &=&  {\cal E}\psi(x, {\cal E}), \,\,\,\,\,\,\,x>\lambda  \nonumber \\
 \left[x\psi'(x,{\cal E})/\psi(x,{\cal E})\right]_{x=\lambda} &=& g(\lambda,{\cal E})
 \label{II} \end{eqnarray}
Certainly, the solutions of eq.(\ref{I}) and (\ref{II}) are identical for $x>{\rm Max}(1,\lambda)$.  One can relate $g(1,{\cal E})$ to $g(\lambda,{\cal E})$ by integrating the Schrodinger equation, which can be written as
\begin{equation}
x\frac{\partial g(x,{\cal E})}{\partial x} = x^2 (U(x) - {\cal E}) -(d-2)g(x,{\cal E}) - g(x,{\cal E})^2
\label{g} .
\end{equation}

\noindent
{\bf Renormalization:}
 Our renormalization group is constructed by recasting the problem in eq.(\ref{II}) into the form of eq.(\ref{I}) so that the boundary condition is always impose at $x=1$.   We define scaled variables
 \begin{equation}
\begin{array}{rclcrcl}
{\cal E}_{\lambda}&\equiv& \lambda^2 {\cal E},&&
 \psi_{\lambda}(x, {\cal E}_{\lambda})& \equiv & \psi(\lambda x,E), \\
 g_{\lambda}(x, {\cal E}_{\lambda})
&\equiv&  g(\lambda x, {\cal E}),      &&
  U_{\lambda}(x) &\equiv&   \lambda^2 U(\lambda x),  \label{glambda} 
\end{array}
 \end{equation}
so that eq.(\ref{II}) becomes
 \begin{eqnarray}
[ -\nabla^2 + U_{\lambda}(x)]\psi_{\lambda}(x, {\cal E}_{\lambda}) &=& {\cal E}_{\lambda}\psi_{\lambda}(x, {\cal E}_{\lambda}), \,\,\,\,\,\,\, x> 1 \nonumber \\
  \left[x\psi'_{\lambda}(x,{\cal E}_{\lambda})/\psi_{\lambda}(x,{\cal E}_{\lambda})\right]_{x=1}&=& g_{\lambda}(1, {\cal E}_{\lambda}),
\label{new1}  \end{eqnarray}
which has the same form as eq.(\ref{I}).   
%Our goal is to classify the solutions of the Schrodinger equation
%by working out the large $\lambda$ behavior of the interactions $U_{\lambda}(r)$ and %$g_{\lambda}(1,{\cal E}_{\lambda})$. 
The ``coupling constants" which parameterize our theory are the potential $U_\lambda(x)$ and the  boundary condition $g_\lambda(1,{\cal E}_\lambda)$, which obey ``flow equations"
\begin{eqnarray}
\lambda \partial_{\lambda} U_{\lambda}(x) & =& x\partial_{x} U_{\lambda}(x)+2U_{\lambda}(x)
\label{rg2} \\
\lambda {\partial_\lambda g_{\lambda}(x, {\cal E}_{\lambda})}
&=& x^2(U_{\lambda}(x) - {\cal E}_{\lambda}) \\\nonumber
&&\,\,- (d-2) g_{\lambda}(x, {\cal E}_{\lambda})- g^{2}_{\lambda}(x, {\cal E}_{\lambda}),
\label{rg1}
\end{eqnarray}
where the derivative of $\lambda$ in $\partial_{\lambda}  g_{\lambda}(x, {\cal E}_{\lambda})$ acts on both subscript  $\lambda$ and the argument ${\cal E}_{\lambda}$.  We can classify various potentials and boundary conditions by their behavior as $\lambda$ increases.

\noindent
{\bf  Flow of $\mathbf{ U_\lambda(x)}$:}  To solve eq. (\ref{rg2}), we expand 
$U_{\lambda}(x)$ in a Laurant series
\begin{equation}\label{uexp}
U_{\lambda} (x)= \sum_{n=0, \pm 1, \pm2, ..} U_{m}(\lambda)  x^{m},
\end{equation}
 where 
$U_{m}(\lambda=1)\equiv U_{m}$ is determined by the initial potential $U(x)$. 
We can explicitly integrate (\ref{rg2}) to find $U_m(\lambda)=U_m \lambda^{m+2}$.  We can derive this same result directly from the definition of $U_\lambda$ in (\ref{glambda}).

Each variable $U_m$ with $m\neq-2$, has three fixed points: $(0,+\infty,-\infty)$ which are (unstable,stable,stable) for $m>-2$, corresponding to ($U_{m}=0$, $U_{m}>0$, $U_{m}<0$); 
and are (stable, unstable, unstable) for $m<-2$, corresponding to ($U_{m}=0$, $U_{m} \rightarrow +\infty$, $U_{m}\rightarrow  -\infty$). The $m=-2$ case corresponds to the inverse square potential $u/x^2$ which does not flow; {\em i.e.}  has a  fixed line  with $U_{m=-2}(\lambda) = u$.

\noindent
{\bf Flow of $\mathbf{g_{\lambda}}$:}
Equation (\ref{new1}) only involves $g_\lambda(x,{\cal E}_\lambda)$ at $x=1$.  Therefore
we write $g_{\lambda}({\cal E}_{\lambda})\equiv g_{\lambda}(1,{\cal E}_{\lambda})$.  Baring pathologies, $g_{\lambda}({\cal E}_{\lambda})$ is a smooth function of ${\cal E}_\lambda$, and can be expanded in a Taylor series
\begin{equation}
g_{\lambda}({\cal E}_{\lambda})=\textstyle \sum_{\ell=0,1,2, ..}G_{\ell}(\lambda){\cal E}_{\lambda}^{\ell}. 
\end{equation}
The function $g_{\lambda}({\cal E}_{\lambda})$ is then characterized by the vector 
${\bf G}(\lambda)  = (G_{0}(\lambda), G_{1}(\lambda), G_{2}(\lambda), ...)$, whose components obey 
\begin{eqnarray}\nonumber
\lambda\partial_\lambda G_0(\lambda)&=&
\left({\textstyle \sum_m\! U_m(\lambda)}\right)-(d-2)G_0(\lambda)-G_0(\lambda)^2\\
\label{newrg}
\lambda\partial_\lambda G_1(\lambda)&=&-1-d G_1(\lambda)-2 G_0(\lambda)G_1(\lambda)\\\nonumber
\lambda \partial_\lambda G_\ell(\lambda)&=& -(d+2\ell-2)G_\ell(\lambda)
-{\textstyle \sum_{\ell^\prime}} G_{\ell^\prime}(\lambda) G_{\ell-\ell^\prime}(\lambda).
\label{flow} \end{eqnarray}
The last expression holds for $\ell>1$, and the sum runs from $\ell^\prime=0$ to $\ell$.
The fixed point solution ${\bf G}^{\ast}$ 
is completely specified by the component $G_{0}^{\ast}$, 
\begin{eqnarray}
G_{1}^{\ast}  &=& -(d + 2G_{0}^{\ast})^{-1}
\label{G1ast} \\
G_{\ell}^{\ast}  &=& -\frac{ \sum_{\ell'=1}^{\ell-1} G_{\ell'}^{\ast}G_{\ell-\ell'}^{\ast}}
{d-2 + 2\ell + 2G_{0}^{\ast} } \,\,\,\,\,\,\,\,\,\, \ell\geq 2
\label{Gellast} \end{eqnarray} 
We can therefore label the behavior of ${\bf G}^{\ast}$ by that of $G_{0}^{\ast}$, which is given by
\begin{equation}
0 = \left(\textstyle\sum_{m}U_{m}^{\ast}\right) - (d-2) G_{0}^{\ast} - G_{0}^{\ast 2}.
\label{G0ast} \end{equation}

\noindent
{\bf  Flow of ${\bf G}$ for non-scale-invariant systems:} 
We first consider potentials that do not have 
a scale invariant component, (i.e. $m=-2$ term is absent in eq.(\ref{uexp})).   There are then three possibilities for the sum $\sum_{m}U_{m}^{\ast}= 0,+\infty,-\infty$.  If the sum is $-\infty$, then (\ref{G0ast}) has no real solution, and there is no fixed point.  We will return to this case later.  If the sum is $+\infty$, then $G_{0}^{\ast}=\pm\infty$, (corresponding to a stable/unstable fixed point) and $G_{j>0}^{\ast}=0$.  Physically, this corresponds to a strongly repulsive long-range potential.

\begin{figure*}
\includegraphics[width=\textwidth]{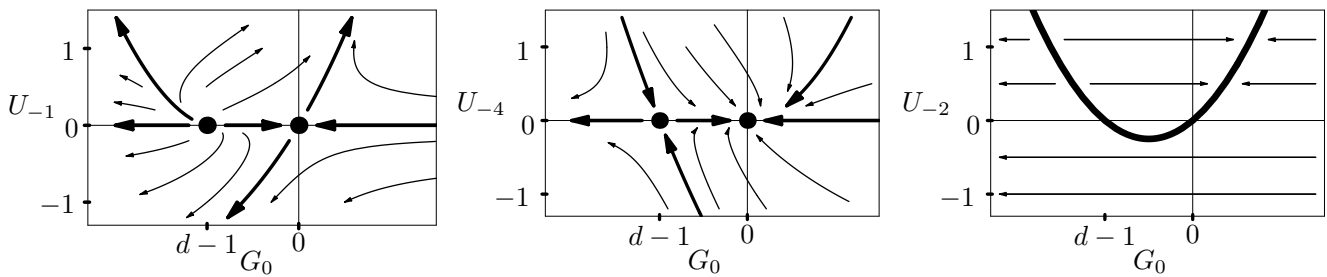}
\caption{(wide) Cross-sections of renormalization group flows.  In each panel, two coupling constants: $U_m$ and $G_0$ are shown.  The distinct structures seen when $m>-2$, $m<-2$ and $m=-2$ are illustrated by the three panels. The curves are calculated for $d=3$ dimensions.  Separatrices are shown in bold \cite{foot}.
}
\end{figure*}

For $G_{0}^{\ast}$ to be finite,  we must take the $U_{m}^{\ast}=0$ fixed point of the potential.   The condition that $U_\lambda(x)\to 0$ as $\lambda\to\infty$ defines a short range potential, and includes potentials which vanish beyond a given radius.  In this case, for $d\neq 2$ we have
\begin{equation}
G_{0}^{\ast}=0\,\,\,\,\,\, {\rm and} \,\,\,\,\,\,2-d,      \,\,\,\,\,\,\,\,\, {\rm for} \,\,\,\,\,\,\,\,\, d\neq 2,
\end{equation} 
with the larger of the two being the stable fixed point.  (See the left and center panel of figure 1). 
The fixed point vector ${\bf G}^{\ast}$ corresponding to each $G_{0}^{\ast}$ can then be obtained from eq.(\ref{G1ast}) to (\ref{Gellast})), with stability identical to that of their 
$G_{0}^{\ast}$ components.  As $d\to2$ these two fixed points coalesce into one: $G_{0}^{\ast}= 0$, which has a saddle-point structure. 
%More explicitly, eq.(\ref{flow}) implies that at $d=2$, $G={0}(\lambda) = 
%1/{\rm ln}(\lambda/c)$, where ${\rm ln}(c) = -1/G_{0}(1)$.  

To understand the meaning of the fixed point solutions $G_{0}^{\ast}$, recall the 
asymptotic form of the wavefunction at low energy and large distances \cite{familiar}: 
\begin{equation}
\begin{array}{rclcl}
\psi(r) &\sim& (k a)^{2-d}-(k r)^{2-d}&\quad& d\neq2\\
&\sim&\ln(r/a)&\quad& d=2,
\end{array}\label{asympt}
\end{equation}
where $E=k^2/2m$ and $a$ is the s-wave scattering length.  Thus, for $d\neq2$, eq.(\ref{asympt}) implies that for large $\lambda$, $G_0=(2-d)a^{d-2}/((\lambda r_0)^{d-2}+a^{d-2})$.  
When $d>2$ ($<2$) and $a\neq\infty$, $G_0$ flows to $0$ ($2-d$), corresponding to weak (strong) coupling.  The
 unstable fixed point at $G_0=2-d$ ($0$) is equivalent to $a=\infty$, corresponding to strong (weak) coupling.  For $d=2$, equation (\ref{asympt}) implies that for large $\lambda$, $G_0=1/\ln(\lambda r_0/a)$. 
 These asymptotics  can also be determined directly from eq.(\ref{newrg}).

\noindent
{\bf Flow of ${\bf G}$ for scale-invariant systems}: 
We now consider inverse square potentials, $U(x)=u/x^2$.  As previously shown, $U_{m=-2}(\lambda)$ has a fixed line solution such that $U_{m=-2}(\lambda) =u$. 
The first equation of (\ref{newrg}) then becomes 
\begin{eqnarray}
\lambda \partial_{\lambda} G_{0}(\lambda) &=&  u - (d-2)G_{0}(\lambda) 
- G_{0}(\lambda)^2 \nonumber \\
  & = & - [ G_{0}(\lambda) -A_{+}][  G_{0}(\lambda) - A_{-}]
\label{G0-2} \end{eqnarray}
where $A_{\pm} = \left( 2-d \pm \sqrt{   (2-d)^2 + 4u} \right)/2$. For 
$u>- (d-2)^2/4$, the flows resemble those of short ranged interactions. There are two 
fixed points, at $G_{0}^{\ast} = A_{\pm}$, with the larger one, $A_{+}$, being the stable fixed point. 
(See the right panel of figure 1. The interaction $U_{2}(\lambda)$ does not flow as mentioned before). 

For a sufficiently attractive inverse square potential, $u<-(d-2)^2/4$, $A_{\pm}$ turns into  a complex
conjugate pair,  $A_{-}= A^{\ast}_{+}$, 
\begin{equation}
A_{+} = \frac{2-d}{2}  +  i  \sqrt{ |u|- \left(\frac{2-d}{2}\right)^2}  \equiv  \alpha + i\beta. 
\end{equation}
As a result, eq.(\ref{G0-2}) has no fixed point solution. Instead, its solution is a periodic function of ${\rm ln}(\lambda)$, as seen by integrating eq. (\ref{G0-2}) to give
\begin{equation}
G_{0}(\lambda) = \alpha + \beta{\rm cot}\left( \beta {\rm ln}\lambda + \theta\right).
\label{periodic} \end{equation}
It then follows from eq.(\ref{newrg}) 
that ${\bf G}(\lambda)$ is also periodic in  ${\rm ln}(\lambda)$, with the properties 
\begin{equation}
{\bf G}(\lambda)= {\bf G}(\lambda \lambda_{o}), \,\,\,\,\,\,\,\,\,\,\,\, \lambda_{o} = 2\pi/\beta
\end{equation}
which in turn implies that 
\begin{equation}
g_{\lambda}(1, {\cal E}) = g_{\lambda\lambda_{o}}(1, {\cal E}). 
\label{gp} \end{equation}
As $\lambda$ increases, $G_{0}(\lambda)$ goes through $\pm \infty$. This behavior can be 
rewritten in a more conventional form of a limit cycle by performing a 
conformal transformation which maps the real axis onto a unit circle in the complex plane,
\begin{equation}\label{phidef}
h(\lambda)=e^{i\Phi}
= \frac{G_0(\lambda)-A_+}{G_0(\lambda)-A_+^*} 
\label{conformal} \end{equation}
The phase is $\Phi=\Phi_0-2 \beta\ln(\lambda)$, where, for a pure inverse square potential, $\tan(2\Phi_0)=\beta/(G_0(1)-\alpha)$.  
The new variable $h(\lambda)$ executes simple periodic motion in the complex plane.
%\begin{equation}
% e^{-2i\beta {\rm ln} \lambda}e^{-2i\theta} = \frac{G_{0}(\lambda) - A_{+}}{G_{0}(\lambda) - A_{+}^{\ast}}  \equiv h(\lambda) 
%  \label{h} \end{equation}
%where the parameters $\theta$  and $\beta$ are determined by the initial condition $G_{0}(1)$ through the relations 
%\begin{equation}
%e^{-2i\theta} = \frac{G_{0}(1)-\alpha - i\beta}{(G_{0}(1)-\alpha+i\beta}, 
%\,\,\,\,\,\,\, {\rm or} \,\,\,\,\,\, {\rm tan}\theta = \frac{\beta}{G_{0}(1) - \alpha}. 
%\end{equation}
%We then have 
%\begin{equation}
%G_{0}(\lambda) = \alpha + \beta{\rm cot}\left( \beta {\rm ln}\lambda + \theta\right).
%\label{periodic} \end{equation}
%which is periodic in ${\rm ln}\lambda$. It then follows from eq.(\ref{G1}) and (\ref{Gell}) 
%that ${\bf G}(\lambda)$ is also periodic in  ${\rm ln}\lambda$, with the properties 
%\begin{equation}
%{\bf G}(\lambda)= {\bf G}(\lambda \lambda_{o}), \,\,\,\,\,\,\,\,\,\,\,\, \lambda_{o} = 2\pi/\beta
%\end{equation}
%which in turn implies that 
%\begin{equation}
%g_{\lambda}(1, {\cal E}) = g_{\lambda\lambda_{o}}(1, {\cal E}) 
%\label{gp} \end{equation}
%It is also interesting to note that quantity $h$ on the right hand sides  of eq.(\ref{h}) is a conformal 
%transformation of $G_{0}$ (if it is extended into the complex plane).  The limit cycle behavior is 
%most explicit in this transformed variable, which executes a periodic motion on a unit circle in the complex plane in time $t={\rm ln}\lambda$  with angular frequency $2\beta$. 
Using the variable $\Phi$, the flow of the entire set of interaction constants is shown schematically in figure 2. 
 
\begin{figure}
\includegraphics[width=\columnwidth]{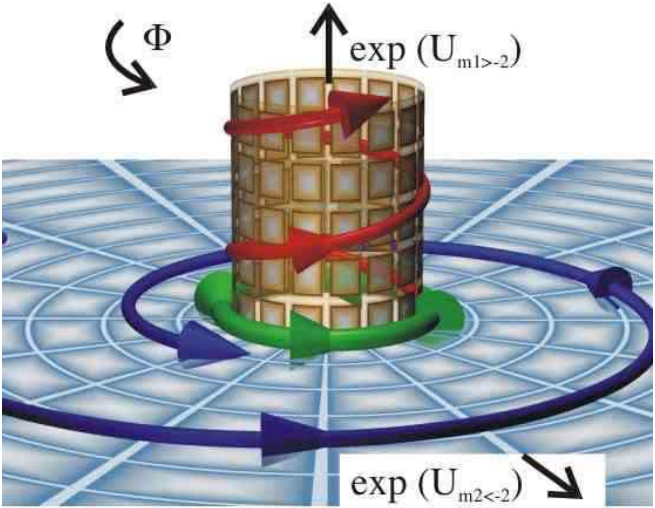}
\caption{ (color) 
Schematic renormalization group (RG) flows near limit cycle.  We depict three of the infinitely many dimensions of coupling constant space.  In cyclindrical coordinates the ($z$,$\rho$,$\phi$) axes represent $\exp(U_{m_1})$, $\exp(U_{m_2})$, and $\Phi$,
where $m_1>-2$ and $m_2<-2$.  These variables, defined in equations (\ref{uexp}) and (\ref{phidef}), are respectively relevant, irrelevant, and cyclic.
%vertical ($\hat z$) axis represents the relevant variable $\exp(U_{m_1})$ with $m_1>-2$.  The radial coordinate ($\rho$) represents the irrelevant variable $\exp(U_{m_2})$ with $m_2<-2$.  The angular coordinate ($\phi$) represents the cyclic variable $\Phi=\arctan(G_0)$.  
The blue horizontal plane and orange vertical cylinder correspond to the surface $U_{m_1}=0$, and $U_{m_2}=0$.  Three RG flows are shown: the blue (red) trajectory with
 $U_{m_1}=0$ ($\neq 0$) and $U_{m_2}\neq 0$ ($=0$) spirals in to (away from) the limit cycle;  the green trajectory lies on the limit cycle with $U_{m_1}=U_{m_2}=0$.}
\end{figure}

\noindent
{\bf Spectral features:}  We now show that this limit cycle leads to a geometric series of bound states and a log-periodic structure in the density of scattering states. 
We begin by noting that the eigenenergies $\{{\cal E_\lambda}^{(n)}\}$ of the Schrodinger equation in eq.~(\ref{new1}) are related to the original eigenvalues  $\{ {\cal E}^{(n)}\}$  in eq.(\ref{I}) as ${\cal E}^{(n)}= {\cal E}_{\lambda}^{(n)} /\lambda^2$.  

For a sufficiently attractive scale inverse square potential $U(x) = -|u|/x^2$, equation (\ref{periodic}) holds,  so that when  the system is scaled by $\lambda=\lambda_0$, the equations of motion and boundary condition in (\ref{new1}) are unchanged.  Therefore the set of energies $\{ {\cal E}_{\lambda_0}^{(n)}\}=\{\lambda_0^2 {\cal E}^{(n)}\}$ must coincide with the original energies $\{ {\cal E}^{(m)}\}$, and each energy $E^{(n)}$ belongs to a geometric series:  for each integer $\nu$, $E^{(n)}/\lambda_0^{2\nu}\in\{ E^{(m)}\}$.  In particular, if a single bound state exists, then there must be an infinite number of bound state whose energies accumulate at threshold.  Additionally, the density of s-wave states $\rho_s(E)=\sum_i \delta(E-E_i)$ must display discrete scale invarience: $\rho_s(\lambda_0^2 E)=\rho_s(E)/\lambda_0^2$, so that $E\rho(E)$ is periodic in $\ln(E)$.  Coarse-graining over these oscillations,  $\rho_s(E)\sim1/E$ as $E\to0$.  This can be 
contrasted with short-range potentials where $\rho_s(E)\sim1/\sqrt{E}$ \cite{add}.

\noindent
{\bf Strongly attractive long range potentials:}
We now revisit the case where $\sum_{m}U_{m}^{\ast} \to-\infty$ as $\lambda\to \infty$.  Let us suppose the Laurant series in (\ref{uexp}) terminates at $\bar m$, and let and $n=\bar m + 2>0$.  For large $\lambda$, $G_0$ obeys
\begin{equation}\label{sat}
\lambda\partial_\lambda G_0(\lambda) = -\gamma \lambda^n - (d-2) G_0(\lambda) - G_0(\lambda)^2,
\end{equation}
 with $\gamma=-U_{\bar m} (1)>0$.  We have neglected terms which are small compared to $\lambda^n$.  As with the case of scale invariant potentials, there exists no stationary solution to these equations and one expects to find periodic behavior instead.  Although (\ref{sat}) is readily integrated in terms of Bessel functions,
the best insight into the behavior of $G_0$ comes from noting that $G_0(\lambda)$ varies more rapidly than $\lambda^n$.  Thus we perform the conformal transformation in (\ref{phidef}), with $A_+(\lambda)=\alpha+i\beta=(1-d/2)+i\sqrt{\gamma\lambda^n - (1-d/2)^2}$, finding 
\begin{eqnarray}\nonumber
\lambda\partial_\lambda h(\lambda) &=& -2 i \beta h+[(h^2-1)/2]\lambda\partial_\lambda {\rm ln} \beta\\ 
&=&
-2 i \gamma^{1/2}\lambda^{n/2} h(\lambda)+{\cal{O}}(\lambda^{n/2-1}).
\end{eqnarray}
The phase therefore behaves as $\Phi = \Phi_0-(4\sqrt{\gamma}/n)\lambda^{n/2}$, i.e. it winds around with an increasing rate as $\lambda$ increases.  This changing period is associated with the fact that 
$\sum_{m}U_{m}(\lambda),$ continuously increases (see the red curve in figure 2), and means that 
despite the cyclic behavior of $\Phi$, this is not a limit cycle.

\noindent
{\bf Ultraviolet properties:}
If the potential is known to arbitrarily short distances, then the scaling equations (\ref{newrg}) can be integrated to $\lambda = 0$.  Instead of low energy physics, one is then exploring high energy physics at short distances. The fixed point solutions at short distances (which are different from the low energy ones) can be obtained similarly and will not be discussed here. We only remark that the scale invarient potential $U=u/r^2$ has an ultraviolet limit cycle when $u<-(d-2)^2/4$.

 This work is supported by NASA GRANT-NAG8-1765  and NSF Grant DMR-0109255.
 
 {\em Notes Added:} After preparing this paper, we discovered that the mathematical arguments given here, including the derivation of limit cycle behavior, have previously been published by Eugene Kolomeisky and Joseph Straley in the context of line-depinning transitions \cite{kolomeisky}.  Due to the large overlap between these works, our paper is not under consideration for publication in any journal.


\begin{thebibliography}{99}
\bibitem{Wilson71} K.G. Wilson, Phys. Rev. D{\bf 3}, 1818 (1971).
\bibitem{Huse}  David A.  Huse, Phys. Rev. B {\bf 24}, 5180 (1981).  The limit cycle in Huse's work may be viewed as an artifact of his decimation scheme, and gives rise to a different class of phenomena than those discussed here.
\bibitem{GW} S.D. Glazek and K.G. Wilson, Phys. Rev. Lett. {\bf 89}, 230401 (2002);
 cond-mat/0303297 (2003).
\bibitem{leclair} A. Leclair, J. M. Roman, and G. Sierra, Nucl. Phys. B {\bf 675}, 584 (2003); hep-th/0312141 (2003); hep-th/0211338 (2002).
\bibitem{Albe-81}
S.~Albeverio, R.~Hoegh-Krohn, and T.S.~Wu,
%``A Class of Exactly Solvable Three-body Quantum Mechanical Problems
%	and the Universal Low-energy Behavior,''
	Phys.\ Lett.\ {\bf 83A}, 105 (1981).
\bibitem{Efimov}  V. Efimov, Phys. Lett. {\bf 33B}, 563 (1970).
%\bibitem{LeClair}  A. LeClair, J. M. Roman, G. Sierra, hep-th/0301042. cond-mat/0211338.
\bibitem{Braaten}  E. Braaten and H.W.. Hammer, Phys. Rev. Lett. {\bf 91}, 102002 (2003).
\bibitem{BEC} 
M. H. Anderson et al. Science {\bf 269}, 198 (1995); C.C. Bradley et al. Phys. Rev. Lett. {\bf 75}, 1687 (1995); K. B. Davis et al. Phys. Rev. Lett. {\bf 75}, 3969 (1995).
\bibitem{Ketterle} 
S. Inouye et al. Nature {\bf 392}, 151 (1998).
%Ketterle's first experiment on Feshbach resonance
\bibitem{barford} 
M. C. Birse, et al.
%J. A. McGovern, and K. G. Richardson, 
Phys. Lett. B {\bf 464}, 169 (1999);
T. Barford and M. C. Birse, Phys. Rev. C {\bf 67}, 064006 (2003).
% "Renormalization group approach to two-body scattering in the presence of long range forces":
% regularizes potential with a "delta-function sphere"
\bibitem{beane}
S. R. Beane et al.
%B. F. Bedaque, L. Childress, A. Kryjevski, J. McGuire, and U. van Kolck,
Phys. Rev. A {\bf 64}, 042103 (2001).
% "Singular potentials and limit cycles"  Uses a square well regulizer
\bibitem{bawin}
M. Bawin and S. A. Coon, Phys. Rev. A {\bf 67}, 042712 (2003).
% "Singular inverse square potential, limit cycles and self-adjoint extensions"
%  Uses beane's regularization, studies limit cycles
%\bibitem{kolomeisky}
%Eugene B. Kolomeisky and Joseph P. Straley, Phys. Rev. B {\bf 46}, 12664 (1992).
\bibitem{qmtext} L. D. Landau and E. M. Lifshitz, {\em Quantum Mechanics,} (Pergamon Press Ltd. London 1958).
\bibitem{desire} This division is also often necessitated by ignorance of the short range structure of $V(r)$.
%\bibitem{general}
%The in general $d$-dimensions ($d\neq 2$) it is convenient to define the scattering length from the $r\to0$ limit of the wavefunction,
%phase shift is defined by the
%large $r$ asymptotic form of the wavefunction
%$\psi(r)\sim (kr)^{(1-d)/2} \cos\left(k r+\delta +(d-1)\pi/4\right)$, where $E=k^2/M$.  
%It is convenient to define the scattering length by the $k\to0$ limit,
%$\cot(\delta)=\cot(\pi(d/2-1))+\frac{\Gamma(d/2)}{2^{2-d}\Gamma(2-d/2)\sin(\pi(1-d/2))} \sgn( a)|k a|^{2-d}$.  
%$\psi(r)\sim \sgn(a) |q a|^{2-d}-|qr|^{2-d}$, so that $G_0=(2-d)/(1+\sgn(a)|a/r|^{2-d})$, where $r=\lambda r_0$.  For $d>2$ ($<2$), if $a\neq\infty$, $G_0$ flows to $0$ ($2-d$), corresponding to weak (strong) coupling.  The
% unstable fixed point at $G_0=2-d$ ($0$) requires $a=\infty$ corresponding to strong (weak) coupling.
\bibitem{familiar}  For $d=1$ and $d=3$, the low energy asymptotic wavefunction is $\psi(r)=\sin(k(r-a))$
%\to -k (a-r)$ 
and $\psi(r)=\sin(k(r-a))/r$
%\to -ka \left(a^{-1}-r^{-1} \right) $.  
%Unlike in $3D$, the non-interacting system in $1D$ has $a=\pi/k\to\infty$; so $a\to0$ is strong coupling %and $a\to\infty$ is weak coupling.  
\bibitem{add} The full density of state will have contributions from all partial waves. 
%\bibitem{correction} From eq.(\ref{conformal}), we have $G_{0} -A_{+} = 
%(A_{+} -  A_{+}^{\ast}) h/(1-h)$. Hence $|G_{0}-A_{+}|^2$ is of order of  $ \gamma \lambda^{n}$.
\bibitem{foot}{The separatrices near the saddle points are determined by linearizing the RG equations, while near the nodes they are determined by the asymptotic large $\lambda$ behavior.  For example, in the leftmost panel of figure 1, a separatrix extends towards the upper-left corner.  All paths to the left of this line reach $G_0=-\infty$ at finite $\lambda$, while those  to the right always have $G_0$ bounded from below.}
\bibitem{kolomeisky}
Eugene B. Kolomeisky and Joseph P. Straley, Phys. Rev. B {\bf 46}, 12664 (1992).
\end{thebibliography}
\end{document}